\documentclass[conference]{IEEEtran}
\usepackage[utf8]{inputenc}
\usepackage[T1]{fontenc}
\usepackage[english]{babel}

\usepackage{xcolor}
\usepackage{hyperref}
\addto\extrasenglish{
    
}

\usepackage{xspace}
\usepackage{graphicx}
\usepackage{balance}
\usepackage{svg}
\usepackage{paralist}
\usepackage{mdframed}
\usepackage{color}
\usepackage{caption}
\usepackage{dingbat}
\usepackage[flushleft]{threeparttable}
\usepackage{colortbl}
\usepackage[]{xcolor}
\usepackage[inline]{enumitem}

\usepackage{subcaption}
\usepackage{listings}

\usepackage{dingbat}

\usepackage{siunitx}

\usepackage[]{xcolor}
\newcommand{\TODO}[1]{\textcolor{red}{#1}\GenericWarning{}{LaTeX Warning: TODO: #1}}\newcommand\todo\TODO

\lstdefinestyle{boxed}{
, numbers=left%
, firstnumber=auto%
, numberblanklines=true%
, frame=trbL%
, numberstyle=\tiny%
, frame=leftline%
, numbersep=7pt%
, framesep=5pt%
, framerule=10pt%
, xleftmargin=15pt%
, backgroundcolor=\color[gray]{0.97}%
, rulecolor=\color[gray]{0.90}%
}

\newcommand{\flacoco}{\textsc{flacoco}\xspace}
\newcommand{\jacoco}{\textsc{JaCoCo}\xspace}
\newcommand{\gzoltar}{GZoltar}
\newcommand{\flacocobot}{\textsc{flacocobot}\xspace}
\newcommand{\github}{{GitHub}\xspace}

\newcommand{\totalProjects}{{10}\xspace}

\newcommand{\totalPR}{{462}\xspace}

\usepackage{multirow}
\usepackage{booktabs}
\definecolor{ForestGreen}{RGB}{34,139,34}
\usepackage{pifont} 
\newcommand{\pr}{PR\xspace}%
\newcommand{\prs}{PRs\xspace}%

\begin{document}

\title{FLACOCO: Fault Localization for Java based on Industry-grade Coverage}

\author{
    \IEEEauthorblockN{André Silva\IEEEauthorrefmark{1}, Matias Martinez\IEEEauthorrefmark{2}\IEEEauthorrefmark{1}, Benjamin Danglot\IEEEauthorrefmark{3}, Davide Ginelli\IEEEauthorrefmark{4}, and Martin Monperrus\IEEEauthorrefmark{1}}
    \IEEEauthorblockA{\IEEEauthorrefmark{1}KTH Royal Institute of Technology}
    \IEEEauthorblockA{\IEEEauthorrefmark{2}Universitat Politècnica de Catalunya - BarcelonaTech}
    \IEEEauthorblockA{\IEEEauthorrefmark{3}Finsit}
    \IEEEauthorblockA{\IEEEauthorrefmark{4}University of Milano - Bicocca}
}




\maketitle



\begin{abstract}
Fault localization is an essential step in the debugging process.
Spectrum-Based Fault Localization (SBFL) is a well-researched fault localization technique, utilizing code-coverage to predict suspicious lines of code.
However, the maturity of SBFL tools is arguably low.
In this paper, we present \flacoco{}, a new fault localization tool for Java.
The key novelty of \flacoco{} is that it is built on top of one of the most used and reliable coverage libraries for Java, \jacoco{}.
We demonstrate that \flacoco{} is the only fault localization tool working on different versions of Java, incl. the most recent LTS Java 17.
We also show that \flacoco is able to perform fault localization on \totalProjects active and popular open-source projects hosted on \github, further demonstrating its robustness.
\end{abstract}

\section{Introduction}



The ubiquitousness of complex software systems has made software engineering an indispensable part of today's society.
During development, developers might introduce undesired behaviors, or so-called ``bugs'' into programs.
Debugging, the task of locating and removing bugs, has long been known to be tedious and time-consuming~\cite{vessey1985expertise}.
To guide developers in debugging, the scientific community has been prolific in proposing techniques to help them identify the faulty sections of code, a task known as ``fault localization''~\cite{wong2016survey}.
Fault localization techniques produce a ranked list of code components (e.g., methods, lines) according to their suspiciousness values.
Spectrum-Based Fault Localization (SBFL)~\cite{wong2016survey} is a popular technique for fault localization, based on code coverage.
However, per \github popularity, tools for SBFL have difficulties in reaching the industry. We believe that it is due to the maturity and robustness of academic tools.

In this paper, we propose \flacoco{}, a novel fault localization tool for Java, meant to be more industry-ready than the state-of-the-art.
The main goals of \flacoco{} are to:
\begin{inparaenum}[1)]
    \item offer good performance in terms of computation time, and
    \item be reliable over a wide range of Java and Java Virtual Machine versions.
\end{inparaenum}
To reach these goals, our key insight is to build \flacoco{} on top of an industry-grade code coverage library for Java called \jacoco{}.
\jacoco{} has a decade-long existence and a wide adoption by the industrial community.
This design decision allows \flacoco{} to support upcoming Java versions as \jacoco{} evolves to support them.

Moreover, for studying fault localization in the field, we propose \flacocobot{}, a software bot that embeds \flacoco in the Continuous Integration (CI) context.
Our key insight is that executing tests in CI is a widely adopted practice~\cite{duvall2007continuous} and, at the time, the only prerequisite for fault localization.
Basically, this allows \flacocobot{} to act whenever a pull request triggers a failing build.
After running SBFL on the respective build, \flacocobot{} enriches the failing \pr{} by commenting with information regarding the fault localization results produced by \flacoco{} (i.e., the suspicious lines).
To our knowledge, \flacocobot{} is the first ever software bot for fault localization.

\flacoco is thoroughly evaluated as follows.
First, to validate \flacoco, we study the extent to which it supports the main Java versions found in the field, comparing it with a state-of-the-art fault localization tool \gzoltar{}~\cite{gzoltar_eclipse}.
Our results show that \flacoco{} is able to support the main Java versions found in the field, and do so faster and to a greater extent than \gzoltar{}.
Second, to evaluate \flacocobot{}, we set up an experiment and configure the bot to monitor pull requests on \totalProjects active and popular open-source projects.

To sum up, the main contributions of this paper are:
\begin{itemize}
    \item The design of a novel fault localization tool for Java based on the industry-grade coverage library \jacoco{};
    
    \item An implementation of the tool, publicly available on \github, and released on Maven Central: \url{https://github.com/SpoonLabs/flacoco}.
    \item A comprehensive series of experiments with \flacoco and \flacocobot, over 835 real world Java bugs from Defects4J and \totalPR pull requests from large and complex open-source projects, showing that \flacoco is effective and applicable. 
\end{itemize}

This paper is a long extension of a 4-pages tech report published on ArXiv~\cite{silva2021flacoco}.
The paper is organized as follows:
\autoref{sec:background} gives an overview about how fault localization works and provides the motivation behind \flacoco{}.
\autoref{sec:design} describes the design decisions of \flacoco{} and how \flacocobot works.
\autoref{sec:experimental_protocol} presents the research questions and describes the methodology followed for our experiments.
\autoref{sec:results} shows the results of the experiments and provides answers to RQs. \autoref{sec:discussion} highlights the current limitations of \flacoco and future improvements that can be done based on the developers' feedback we received. \autoref{sec:related-work} presents the related work.
\autoref{sec:conclusion} provides final remarks.

\section{Background}
\label{sec:background}

\subsection{Spectrum-Based Fault Localization and Code Instrumentation}

Spectrum-Based Fault Localization is founded on three aspects in order to rank suspicious lines of code:
\begin{inparaenum}[\it i)]
    \item tests results,
    \item code coverage per test cases, and
    \item a formula which computes a suspiciousness score based on the coverage information.
\end{inparaenum}

Code coverage information is, traditionally, collected using code instrumentation.
Code instrumentation consists of inserting code into executable binaries in order to extract information regarding the program's execution.
In the case of code coverage, instrumentation yields which sections of source code are executed.

\subsection{Motivation for \flacoco{}}
\label{sec:motivation}

\begin{figure}[ht!]
    \centering
    \includegraphics[width=0.8\columnwidth]{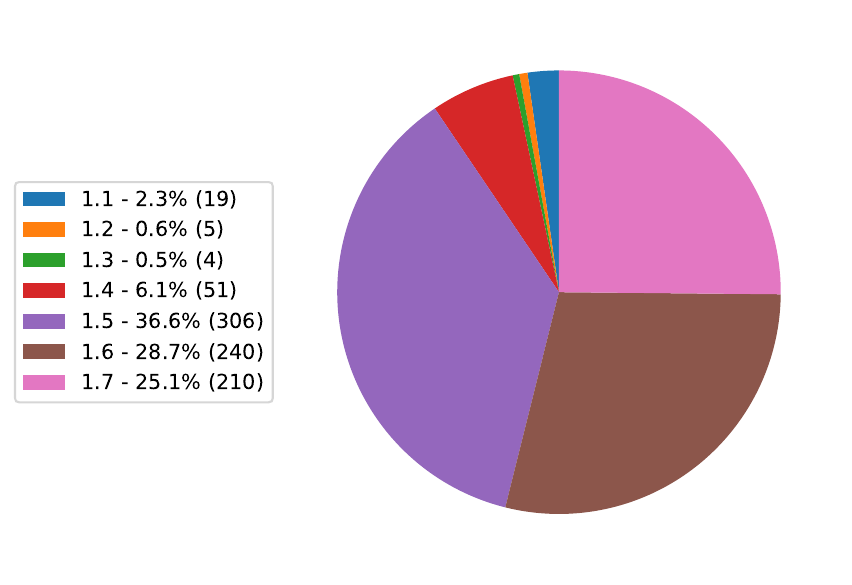}
    \caption{Proportion of Defects4J bugs per bytecode version. A fault localization tool for Java must be able to support all of them.}
    \label{fig:classfiles_two}
\end{figure}

Spectrum-Based Fault Localization relies on code instrumentation to obtain spectra (i.e., coverage and test results information).
However, binary code formats are complex and constantly evolving.
For example, Java has 21 different types of bytecode formats.
In order to perform fault localization at large, one must be able to instrument code on many different code versions.

To substantiate this claim, we study Defects4J V2.0~\cite{just2014defects4j}, a benchmark of 835 Java bugs found in real industrial projects, widely used by the software engineering research community, including evaluation of test generation tools~\cite{Shamshiri2015GeneratedTest} and program repair approaches~\cite{martinez2017D4jRepair}.
Defects4J was intentionally built to run on Java 1.7 and allows bugs to target any bytecode version lower or equal than 1.7 .

We have measured the diversity of bytecode versions in Defects4J.
Defects4J uses 7 different versions, shown in \autoref{fig:classfiles_two}, ranging from 1.1 to 1.7.
The most frequent dominant Java version is 1.5, which is present in 36.6\% of the bugs. 
Furthermore, many bugs blend different versions: $76\%$ include a single Java version, and $24\%$ of the bugs include files of two different versions.
Consequently, to perform spectrum-based fault localization on Defects4J, one needs to be able to obtain spectra for all those versions.

In this paper, our aim is to create an SBFL tool that would work well for different Java versions, incl. the most recent ones.

\section{Design of \flacoco}
\label{sec:design}

In this section, we expose \flacoco{}, a novel fault localization tool for Java.

\subsection{Fault Localization}

\begin{figure}
    \centering
    \includegraphics[width=0.5\columnwidth]{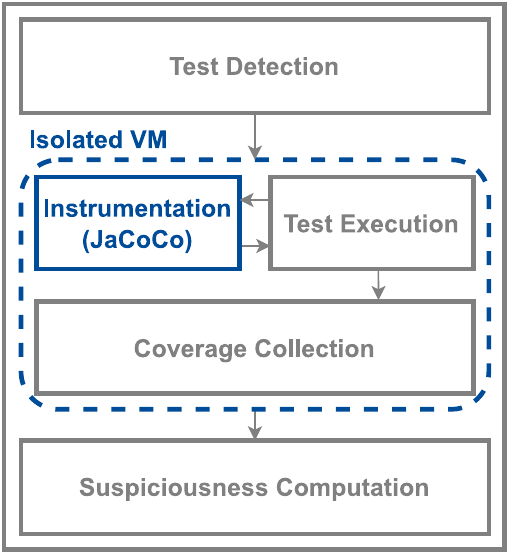}
    \captionsetup{width=2\columnwidth, margin={0\columnwidth, 0\columnwidth}, justification=centering}
    \captionof{figure}{Overview of \flacoco{}. The core novelty and main strength is to be founded on the industry-grade code instrumentation library JaCoCo.}
    \label{fig:overview_core}
\end{figure}

As shown in \autoref{fig:overview_core}, \flacoco consists of 4 steps for fault localization:
\begin{inparaenum}[\it a)]
    \item Test Detection,
    \item Instrumentation \& Test Execution,
    \item Coverage Collection, and
    \item Suspiciousness Computation.
\end{inparaenum}

\subsubsection{Pipeline}

\label{sec:arch}


\emph{Test Detection}
\flacoco{} detects the tests to be executed and analyzed by scanning the compiled classes (Java bytecode) from the project under analysis and filtering those that correspond to test cases according to the test framework's specification.
\flacoco{} supports \textit{JUnit 3}, \textit{JUnit 4}, and \textit{JUnit 5}: this is the only fault localization tool with such a wide test driver support to the best of our knowledge.


\emph{Instrumentation \& Test Execution}
\flacoco{} executes all the previously identified tests in a separate process.
This execution is monitored, via a custom Java agent, in order to  record the lines covered by each test case.
Monitoring does on-the-fly code instrumentation. The instrumentation of the executed classes takes place at class loading time during test execution, and is fully provided by the industry-grade Java code coverage library \jacoco{}~\footnote{\url{https://www.jacoco.org/jacoco/}}.
The instrumentation in \jacoco{} supports all Java bytecode versions, from very old ones (e.g., Java 1.4 was released in Feb 2002) to the latest ones.

\emph{Coverage Collection}
At the end of each completed unit test execution,
\flacoco{} collects and stores the following information:
\begin{inparaenum}[\it a)]
    \item test result (passing, failing),
    \item if the test throws an exception, and
    \item coverage per line for each class executed by the test.
\end{inparaenum}

\emph{Suspiciousness Computation}
At the end of the test suite execution (i.e., after all tests are executed), 
\flacoco{} loads the results from each test execution in order to compute the suspiciousness score of each covered line.
The score of a line is based on the tests that covered it, the results of those tests, and a specified formula. 
The default formula implemented in \flacoco{} is Ochiai \cite{abreu2006evaluation}.

\subsubsection{Main Design Decisions}
\label{sec:designdecision}

\emph{Instrumentation}
Instrumenting for coverage computation is a hard problem \cite{tengeri2016negative}.
In addition, there is a recurring need for updating the instrumentation tools as the bytecode evolves.
To overcome this problem, we decide to rely on an industry-grade, community supported code coverage library.
We build \flacoco{} on top of \jacoco{}, which is actively maintained and updated.
Consequently, \flacoco{} supports from Java 1 to Java 17 bytecode.
Hence, \flacoco{} supports the majority of Java versions and all Java LTS versions.


Another advantage of \jacoco{} is performance. 
\jacoco{} places its probes at certain positions of the control-flow of the program methods.
On the contrary, other tools such as \gzoltar{} use line-based instrumentation, placing probes at the beginning of each source code line, resulting in a larger number of probes.
As demonstrated in our experiments, control-flow based code instrumentation is key for performance. 

\emph{Test Isolation}
\flacoco{} performs test execution with an isolated process.
This strategy has the following major advantages.
\begin{inparaenum}[\it 1)]
    \item Controlling the execution time of the test: \flacoco{} can interrupt a single test execution after a timeout is reached, in a configurable manner. 
    This allows \flacoco{} to avoid non-terminating processes occasioned by, for instance, infinite loops.
    \item Avoiding conflicts between the classpaths of the application under analysis and of \flacoco{}: the process that runs the test only has the dependencies of the application under test, with no spurious dependency shadowing.
    \item Enabling instrumentation at class loading time.
\end{inparaenum}


\emph{Extensibility}
In Spectrum-Based Fault Localization, ranking lines by suspiciousness values can be done according to different formulas such as Ochiai~\cite{abreu2006evaluation}.
\flacoco{} provides a common interface that can be used to integrate other SBFL formulas. 
\flacoco{} is designed to enable incorporation of other methods by providing a common Java fault localization interface, allowing, for example, mutation-based fault localization and machine learning fault localization techniques.

\emph{Support for exception bugs}
\jacoco{} inserts probes at control-flow statements of Java methods, but not at the beginning of each line. 
Lines are marked as covered when a probe placed \emph{after} is marked as executed.
As a result, when an exception is thrown between two probes, some actually covered lines can be missing, as mentioned in the reference documentation.
This is a result of doing efficient control-flow based instrumentation.
However, this affects fault localization for those bugs related to thrown exceptions.
To accommodate this limitation, \flacoco{} parses the stack trace of each exception thrown during test execution to collect the lines which have been executed. 
This is done by parsing the source code of the class each stack trace line belongs to, and identifying the blocks they are in. 
Thus, \flacoco{} is able to recover all lines which belong to that block and have been executed but were not included by \jacoco{} in the coverage result.
This effectively enables \flacoco{} to reconstruct coverage lost due to \jacoco{}'s limitation.
This neat feature is important for research, since academic bug benchmarks tend to contain exception bugs.

\emph{Interfaces}
\flacoco{} provides fault localization through the command line interface (CLI) and through a Java API.
This API allows researchers and developers easy integration of \flacoco in their tools.
For instance, the automated program repair framework Astor~\cite{martinez:hal-01321615} uses \flacoco{} through that API.
Furthermore, \flacoco{} provides, via its API, an annotated AST (abstract syntax tree) with suspiciousness information in order to facilitate the integration of \flacoco{} with tools that work at the AST level.
For that, \flacoco{} creates the ASTs of the applications using Spoon \cite{Pawlak_Spoon_A_Library}.
Then, for each covered line, \flacoco{} searches for the top-most AST node whose position starts and ends at the given line, when possible. 
If not, the line is mapped to the AST node whose position includes the given line and whose amplitude (i.e.\ lines that it covers) is the smallest.
Finally, \flacoco{} assigns the suspiciousness of the covered line to the mapped AST node.


\begin{figure*}
    \includegraphics[width=2\columnwidth]{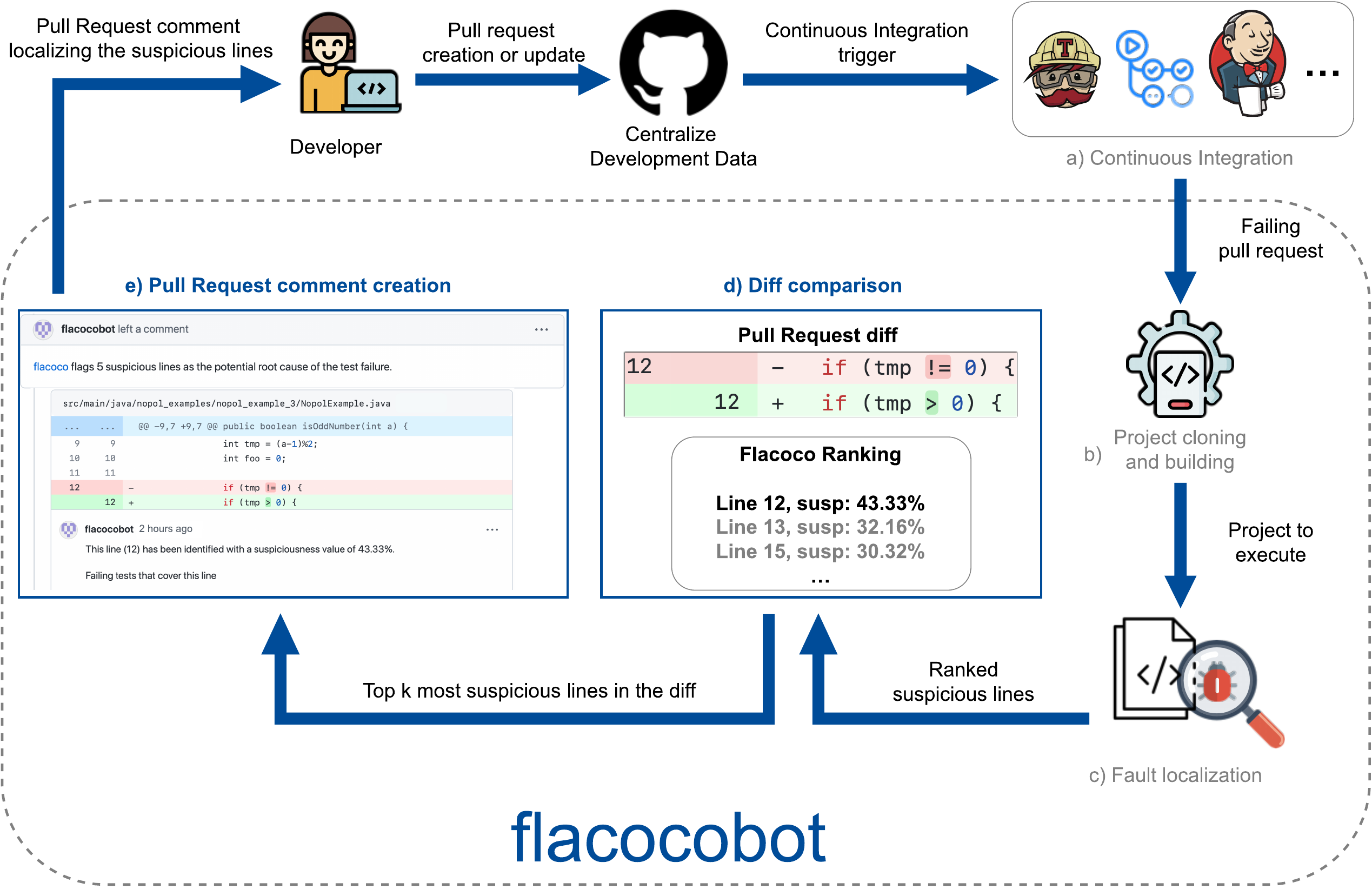}
    \captionsetup{width=2\columnwidth, margin={0\columnwidth, 0\columnwidth}, justification=centering}
    \captionof{figure}{Overview of the integration of \flacoco in a development pipeline in order to perform fault localization in the field.}
    \label{fig:overview:bot}
\end{figure*}

\subsection{Integration in Continuous Integration}

Development workflows based on Continuous Integration (CI) to run test cases are widely adopted in industry.
Since one of our goals is to provide a fault localization tool which is industry-ready, we decide to integrate \flacoco{} in Continuous Integration.
In this way, we can validate our tool in the field on real world projects using CI.
We now present the integration of \flacoco in Continuous Integration, a tool called \flacocobot.
As shown in \autoref{fig:overview:bot},
\flacocobot is characterized by five main conceptual components:
\begin{inparaenum}[\it a)]
    \item Continuous Integration,
    \item Project cloning and building,
    \item Fault localization with \flacoco{},
    \item Diff comparison, and
    \item Pull Request comment creation.
\end{inparaenum}

When developers create or update \prs{} on a centralized development platform (e.g. \github), an event triggers the CI (Step a).
The CI executes the tests to check that the \pr{} does not introduce any regression.
Autonomously, \flacocobot scans the status of the CI from a list of projects.
When \flacocobot detects a failing build triggered by a \pr, it launches a fault localization process to find and rank the suspicious code lines.

The first step clones the project's repository and builds the project in an isolated Docker environment (Step b).
Then, using the compiled program, tests and dependencies, \flacocobot{} computes suspicious lines with a value ranging from 0 (lowest probability of being buggy) to 1 (highest probability) (Step c).

After computing the suspicious lines, \flacocobot{} prepares a message to show the fault localization information to the developer. 
To stay in the developer's current mental context, \flacocobot primarily pinpoints the most suspicious lines that are related to the changes. 
This avoids costly context switches while maximizing the likelihood of being useful.

\flacocobot{} first computes the diff between the main code base and the new version proposed.
From those lines (affected by the \pr), \flacocobot{} selects the $k$ changed lines with the highest suspicious value (Step d).
Finally, \flacocobot creates a comment on the failure \pr, such as the one shown in Figure \ref{fig:overview:bot}, with a message reporting selected suspicious lines (Step e).
For each line, \flacocobot indicates the suspiciousness to be faulty (as a percentage) and the failing test cases that cover the line.
This visualization piggybacks on the \github API to create line-based comments.

\section{Experimental Methodology}
\label{sec:experimental_protocol}

\subsection{Research Questions}

To evaluate \flacocobot{}, we elaborate the following research questions:

\begin{enumerate}[]
    \item[RQ$_1$] (Applicability) To what extent does \flacoco{}'s instrumentation support the main Java versions found in the field?
    
    \item[RQ$_2$] (Field Fault Localization) To what extent is \flacocobot{} capable of performing dynamic analysis and fault localization for \prs{} in the field?
    
\end{enumerate}

\subsection{Methodology for RQ1}

The core idea is to compile each Defects4J bug to different target bytecode versions.
Then, we run fault localization with both \flacoco{} and \gzoltar{} on each pair of $\langle$bug, version$\rangle$.
We compare with \gzoltar{} only because it is, to the best of our knowledge, the only other Java fault localization tool actively maintained and documented.
The experimental pipeline is, therefore, composed of three steps:
\begin{inparaenum}[a)]
    \item compilation of bugs,
    \item execution of fault localization tools, and
    \item analysis of results.
\end{inparaenum}
We now detail them:

\textbf{Compilation of bugs}
We modify the configuration of each bug in order to compile it to specified target versions.
We target Java Long-Term-Support (LTS) bytecode versions (Java 7, Java 8, Java 11, and Java 17)~\footnote{https://www.oracle.com/java/technologies/java-se-support-roadmap.html}.

After the compilation step, we select the bugs whose compiled files 
belong exclusively to the targeted version, since these are the files that are instrumented.
The Java 7 and Java 8 targets were executed, respectively, on OpenJDK 8 and OpenJDK 11 because of infrastructure dependencies.
The Java 11 and Java 17 targets were executed, respectively, on OpenJDK 11 and OpenJDK 17.

\textbf{Execution of tools}
We execute both tools on the selected bugs for each considered version, as well as the default Defects4J configuration as a baseline.
The set of test cases to execute, as provided by Defects4J, is given to each tool.
Each execution is limited by a timeout of 1 hour.
The JVMs used for execution are the same as in the compilation stage.

We evaluate the latest releases to date of both \flacoco{}\footnote{\url{https://github.com/SpoonLabs/flacoco/commit/38e635b0ff811581c0eb7cbb390f92f22faac099} - v1.0.5} and \gzoltar{}\footnote{\url{https://github.com/GZoltar/gzoltar/commit/e93c2aabc0d91d8ddb75acfd46ecb300f18399d8} - v1.7.3}~\cite{gzoltar_eclipse}.

\textbf{Analysis of results}
We verify the generated fault localization results, and consider as valid all non-empty ranked lists of lines with non-zero suspiciousness scores.

In addition to this, we record the execution time of each run.
This is done to compare the efficiency of each tool in terms of execution time.
In particular, we test the hypothesis that \gzoltar{}'s execution times are statistically greater than \flacoco{}'s by performing a one-tailed paired samples t-test for each target ($H_0$: the population mean of differences is zero; $H_1$: the mean of \gzoltar{}'s population is greater than \flacoco{}'s).


\subsection{Methodology for RQ2}
\label{sec:rq2-methodology}

To answer this research question, we search for currently active projects that can be built by \flacocobot, with the aim of analyzing active PRs.
Moreover, we ask developers to give permission to run \flacocobot on their projects in order for the bot to leave comments on failed PRs with suggested suspicious lines.
In this way, we can see if and how developers react to them. 

The process used to answer the research question is thus characterized by four main phases:
\begin{inparaenum}[\it 1)]
\item Sampling of Projects, 
\item Collection of Fault Localization Results, and 
\item Analysis of Fault Localization Results.
\end{inparaenum}

\subsubsection{Sampling of Projects} 
To evaluate the capability of \flacocobot finding and suggesting suspicious lines in real software projects in the field, we select \github repositories of Java projects exploiting two different sources: Bears~\cite{DBLP:journals/corr/abs-1901-06024} and the \github API\footnote{\url{https://docs.github.com/en/rest}}.
Bears~\cite{DBLP:journals/corr/abs-1901-06024} is a benchmark of buggy versions of 72 open-source projects.
The \github API allows us to query \github repositories.

From both sources, we sample projects according to the following criteria:
\begin{enumerate*}
    \item It is possible to build them using \flacoco,
    \item The dominant programming language is Java,
    \item The number of stars is greater than 100,
    \item The last commits on the default branch are not older than one week as of April, 4th, 2022 (for the first phase of the experiment) and not older than one week as of June, 6th, 2022 (for the second phase of the experiment),
    \item There exists at least one open and failed pull request as of April, 4th, 2022 (for the first phase of the experiment), and June, 6th, 2022 (for the second phase of the experiment).
\end{enumerate*}
We save the first 10 projects for further analysis.

\subsubsection{Collection of Fault Localization Results}

We run \flacocobot{} on the 10 projects selected for the experiment, so that when a pull request with a failing CI is detected, it computes and stores the fault localization information.

In particular, we configure \flacocobot{} with:
\begin{inparaenum}[\it a)]
    \item a timeout of 16 minutes for fault localization,
    \item 0.1 as the suspiciousness threshold,
    \item 5 as the maximum number of results to be pushed by \flacocobot{} as PR suggestions, and
    \item OpenJDK 11 as the execution JVM.
\end{inparaenum}

\subsubsection{Analysis of Fault Localization Results}

We expect that \flacocobot{}
\begin{inparaenum}[\it 1)]
\item is capable of applying fault localization on a failing pull request, obtaining a non-empty result, and
\item that the results (i.e., lines with suspiciousness score) include some of the lines modified by the failing pull request.
\end{inparaenum}

To measure the degree of success of \flacocobot{}, we manually analyze each failing pull request with the goal of labeling it with a category according to the outcome.
We start the analysis process with two categories, both corresponding to what we expect from \flacocobot{}, that is, to obtain a non-empty result, and the result includes as suspicious some lines from diff.
Then, while inspecting the pull requests, if any existing category does not describe the performance of \flacocobot{} on a particular pull request, we create a new category.

The analysis of each pull request consists in inspecting:
\begin{inparaenum}[\it a)]
\item the result of the fault localization (either a comment on the pull request or the \flacocobot{}'s logs);
\item if the result is empty (e.g., \flacocobot{} does not find any suspicious line), we inspect the execution log from \flacocobot{} in order to detect whether the failure or malfunction is;
\item if the log does not provide enough information (e.g., no error thrown by \flacocobot), we inspect the logs from the CI job related to the pull request, and we compare it with the \flacocobot{} logs. In particular, we search for the build results and the results associated with the tests executions.
\end{inparaenum}
The categories we found during the analysis are the following.

\paragraph{\texttt{SUSPICIOUS\_LINES\_FOUND\_IN\_PR\_DIFF}}
It represents the successful cases for which \flacocobot{} detects a failed pull request, it is able to perform fault localization and it finds suspicious lines that are contained in the set of lines changed by the pull request.

\paragraph{\texttt{NO\_RELEVANT\_SUSPICIOUS\_LINES\_FOUND}}
It represents the successful cases in which \flacocobot{} detects a failed pull request, performs fault localization, finds failing tests, but does not report any suspicious location, as the suspiciousness values of all lines are below the suspiciousness threshold.

\paragraph{\texttt{SUSPICIOUS\_LINES\_FOUND\_OUT\_OF\_PR\_DIFF}}
It represents the cases in which \flacocobot{} detects a failed pull request, performs fault localization, finds suspicious lines, but these lines are not contained in the set of lines changed by the pull request, which goes beyond the \github{} suggestion capabilities.

\paragraph{\texttt{OUT\_OF\_SCOPE\_CI\_FAILURE}}
It represents the cases for which there is a failure or an error in CI, but it is not related to failing test cases. 
For this reason, the PR is marked as failing on \github, but \flacocobot does not find any test failure.
For example, commit 1a2f370 from \texttt{Apache/Rocketmq} project has a failure related to low test coverage. 
Consequently, this failure is outside of \flacocobot{}'s scope.

\paragraph{\texttt{FLACOCOBOT\_BUILD\_REPRODUCTION\_ERROR}}
It represents the cases for which \flacocobot{} either:
\begin{inparaenum}[a)]
    \item is unable to reproduce the CI build (e.g., due to problems with dependencies, incorrect environment),
    \item is unable to execute the expected failing tests, leading to an empty fault localization result (i.e., 0 suspicious lines), or
    \item is able to execute the expected failing tests, but these return a different result (e.g. flaky tests).
\end{inparaenum}
We recall that \flacocobot{} runs autonomously to the CI of each project, removing the need for developers to change their CI configurations.
Thus, \flacocobot{} trades increased usability for potential build reproduction errors.

\paragraph{\texttt{FAILURE\_HIDDEN\_BY\_FORCE\_PUSH}}
It represents the cases for which the CI is green, even though the pull request has previously been detected as failed. The reason for having this situation is that a new commit has been force pushed. This push removes the commit that produces the failure and updates the previous status of the PR, hiding the failure that triggered \flacocobot{}.

\paragraph{\texttt{PR\_NOT\_FAILED}}
It represents false positives of the CI. 
It could happen that the API used to detect the status of PR returns a failure status, but in fact the pull request is not failed. 
For this reason, \flacocobot{} does not find any failed tests.

\paragraph{\texttt{FLACOCOBOT\_CRASH}}
It represents the cases where \flacocobot{} crashes. In this case, \flacocobot cannot execute fault localization due to an internal limitation, for example, timeout errors and failures during the instrumentation of the programs under analysis.

\section{Experimental Results} 
\label{sec:results}

\subsection{RQ1 (Applicability) To what extent does \flacoco{}'s instrumentation support the main Java versions found in the field?}

Out of the original 835 bugs from Defects4J V2.0, we compute the intersection of 191 bugs that are compilable and executable on all Java LTS versions. We use them in this experiment.
We provide the listings of bugs compilable to each Java LTS version, as well as the listing of the 191 cross-version bugs, as part of the reproduction package.

\begin{table}[!ht]
\centering
\setlength{\extrarowheight}{0pt}
\addtolength{\extrarowheight}{\aboverulesep}
\addtolength{\extrarowheight}{\belowrulesep}
\setlength{\aboverulesep}{0pt}
\setlength{\belowrulesep}{0pt}
\caption{Applicability comparison on the 191 selected Defects4J~\cite{just2014defects4j} bugs.}
\label{tab:rq1}
\begin{tabular}{ccc} 
\toprule
                                                            & \multicolumn{2}{c}{\textbf{Fault Localization}}                 \\ 
\cline{2-3}
                                                            & \textbf{\gzoltar{} v1.7.3}    & \textbf{\flacoco{} v1.0.5}                                    \\ 
\hline
\rowcolor[rgb]{0.949,0.949,0.949} \textbf{Defects4J Config} & 108 (56.5\%) & 125 (65.4\%)                        \\
\textbf{Java 7}                                             & 111 (58.1\%) & 125 (65.4\%)  \\
\rowcolor[rgb]{0.949,0.949,0.949} \textbf{Java 8}           & 110 (57.6\%) & 125 (65.4\%)  \\
\textbf{Java 11}                                            & 121 (63.3\%) & 125 (65.4\%)  \\
\rowcolor[rgb]{0.949,0.949,0.949} \textbf{Java 17}          & 0 (0.0\%)    & \textbf{125 (65.4\%)}  \\ 
\hline
\textbf{Total}                                              & 450 (47.1\%) & \textbf{625 (65.4\%)}  \\
\bottomrule
\end{tabular}
\end{table}

\autoref{tab:rq1} shows the results of executing both \flacoco{} v1.0.5 and \gzoltar{} v1.7.3 on those 191 bugs.
The table is read as follows:
The columns under the label \textbf{Fault Localization} report the number of bugs for which each fault localization tool produces valid fault localization results (i.e., a non-empty ranked lists of lines with non-zero suspiciousness scores), followed by the proportion in parentheses expressed as a percentage over 191 bugs.
It is divided into two columns, one for each fault localization tool: \gzoltar{} v1.7.3 and \flacoco{} v1.0.5.
Each row reports the result for the considered version, while the final row aggregates the results.

In total, \flacoco{} is able to generate a valid fault localization result for 625 pairs of $\langle$bug, version$\rangle$, as opposed to 450 for \gzoltar{}.
For example, \flacoco successfully output fault localization results for 125 bugs compiled to the Java 7 bytecode format.
\flacoco{} maintains its fault localization capability across all versions, whereas \gzoltar{}'s performance varies depending on the target version.

Most importantly, we find that \gzoltar{} cannot perform fault localization for Java 17, the latest LTS version to date.
In contrast, \flacoco{} is capable of dealing with Java 17, obtaining the same performance as when dealing with older Java versions.

We now explain why and when \flacoco{} and \gzoltar{} fail to report valid fault localization results.
After manual analysis, failures in reporting valid fault localization results are found to be due to two distinct reasons:
\begin{inparaenum}[a)]
    \item Execution failures, which can include cases such as Out Of Memory (OOM) errors, timeouts or instrumentation failures;
    \item No failing tests, which includes cases where the considered tool does not report any failing test, thus leading to an empty list of suspicious lines.
\end{inparaenum}
In total, \flacoco{} failed to report valid fault localization results 330 times, always due to execution failures.
On the other hand, \gzoltar{} failed to report valid fault localization results 505 times, where 151 where due to execution failures and 354 due to no failing tests being reported.

\begin{figure}[h!]
    \centering
    \includegraphics[width=0.9\columnwidth]{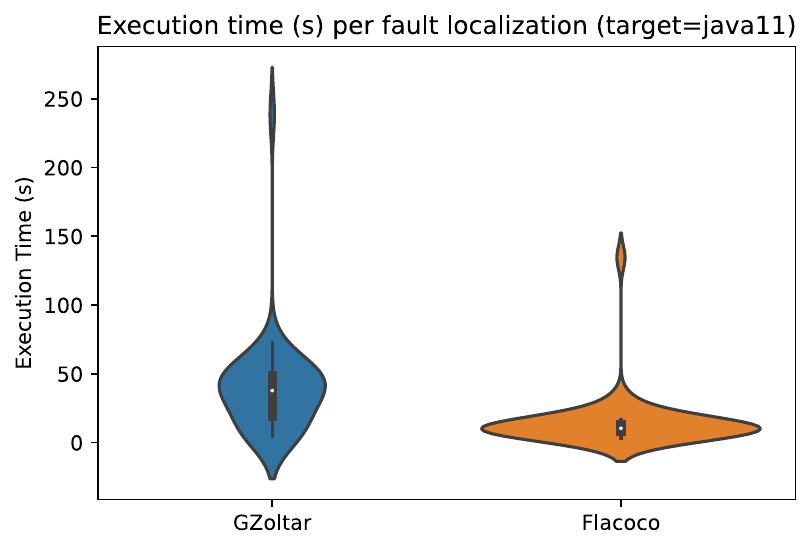}
    \caption{Violin plot of the execution time, in seconds, per fault localization tool for Java11.}
    \label{fig:exec_time_java11}
\end{figure}

We now focus on the comparison of execution times.
We consider, for each version, the bugs for which both tools obtain fault localization results.
\autoref{fig:exec_time_java11} shows the violin plot of the execution time for Java 11, the latest LTS version supported by both \flacoco{} and \gzoltar{}, showing that \flacoco{} is faster than \gzoltar{}.
The remaining violin plots can be found in \flacoco's reproduction package.
After performing a one-tailed paired samples t-test, we find a significant difference ($\alpha=0.05$) for all target versions considered:
\begin{inparaenum}[(i)]
    \item Defects4J Config: $p=\num{8.183e-14}$,
    \item Java 7: $p=\num{1.090e-13}$,
    \item Java 8: $p=\num{1.620e-13}$, and
    \item Java 11: $p=\num{8.131e-18}$.
\end{inparaenum}
Thus, we reject the null hypotheses ($H_0$: the population mean of differences is zero).


\fbox{\begin{minipage}[t]{0.45\textwidth}
\textbf{Answer to RQ$_1$}:
\flacoco{} supports several different Java versions, thanks to being founded on a very-well engineered Java bytecode instrumentation library (\jacoco{}).
\flacoco{} is the only fault localization tool that supports the latest Long-Term-Support (LTS) Java version, Java 17. 
On the considered benchmark, \flacoco{} is better than \gzoltar{} with respect to both effectiveness and performance.
\end{minipage}}

\subsection{RQ2 (Field Fault Localization) To what extent is \flacocobot{} capable of performing dynamic analysis and fault localization for \prs{} in the field?}
\label{sec:rq2-results}

The ten real-world projects selected for the field study are as follows.
Six out of 72 projects of Bears~\cite{DBLP:journals/corr/abs-1901-06024} fulfilled the criteria. We then selected other 4 projects found using the \github API.
\autoref{tab:experiment-projects} shows the list of the projects. 
In particular, for every project, \autoref{tab:experiment-projects} reports the lines of Java code (column \emph{LoC)}, the number of test cases (column \emph{\# Tests}), the number of Maven modules (column \emph{\# Modules}), and the bytecode version of the compiled classes (column \emph{Bytecode Version}).

\begin{table*}[t!]
    \centering
    \setlength{\extrarowheight}{0pt}
    \addtolength{\extrarowheight}{\aboverulesep}
    \addtolength{\extrarowheight}{\belowrulesep}
    \setlength{\aboverulesep}{0pt}
    \setlength{\belowrulesep}{0pt}
    \caption{Projects analyzed by \flacocobot during the experiment (Data related to the status of projects on April, 4th, 2022).}
    \label{tab:experiment-projects}
    \begin{tabular}[t]{lrrrl}
    \toprule
    \textbf{Project}  & \textbf{LoC} & \textbf{\# Tests} & \textbf{\# Modules} & \textbf{Bytecode Version} \\
    \hline
    \rowcolor[rgb]{0.949,0.949,0.949}
    apache/dubbo & $>$ 196k & $>$ 2k & 24 & Java 8 \\
    apache/pinot & $>$ 340k & $>$ 4,7k & 18 & Java 11 \\
    \rowcolor[rgb]{0.949,0.949,0.949}
    apache/rocketmq & $>$ 106k & $>$ 380 & 15 & Java 8 \\
    apache/shardingsphere & $>$ 235k & $>$ 26k & 13 & Java 8 \\
    \rowcolor[rgb]{0.949,0.949,0.949}
    confluentinc/kafka-connect-jdbc & $>$ 18k & $>$ 1,3k & 1 & Java 8 \\
    debezium/debezium & $>$ 169k & $>$ 1,1k & 23 & Java 8 \& 11 \\
    \rowcolor[rgb]{0.949,0.949,0.949}
    jenkinsci/kubernetes-plugin & $>$ 15k & $>$ 240 & 1 & Java 8 \\
    INRIA/spoon & $>$ 152k & $>$ 2k & 1 & Java 11 \\
    \rowcolor[rgb]{0.949,0.949,0.949}
    vert-x3/vertx-jdbc-client & $>$ 7k & 300 & 1 & Java 8 \\
    vert-x3/vertx-web & $>$ 75k & $>$ 1,8k & 11 & Java 8 \\
    \bottomrule
    \end{tabular}
\end{table*}


For all identified projects, we ran \flacocobot for 8 weeks in two different phases of 4 weeks: the first one started on April 4th, 2022, and the second one started on June 6th, 2022.
Overall, \flacocobot analyzed 356 pull requests and managed to perform fault localization and find suspicious locations for 9 out of 10 projects, demonstrating its applicability in the field.

We now present the fault localization results by category shown in \autoref{tab:rq2}, considering the 356 pull requests. We also provide more details on the remaining 106 cases not addressed by the bot.

\begin{table}[!ht]
 \centering
\setlength{\extrarowheight}{0pt}
\addtolength{\extrarowheight}{\aboverulesep}
\addtolength{\extrarowheight}{\belowrulesep}
\setlength{\aboverulesep}{0pt}
\setlength{\belowrulesep}{0pt}
    \caption{Fault Localization results on the PRs analyzed by \flacocobot.}
    \label{tab:rq2}
    \begin{tabular}[t]{lr} 
    \toprule
    \textbf{Category} & \textbf{\# PRs}  \\ 
    \hline
    \rowcolor[rgb]{0.949,0.949,0.949} SUSPICIOUS\_LINES\_FOUND\_IN\_PR\_DIFF & 5 \\
    NO\_RELEVANT\_SUSPICIOUS\_LINES\_FOUND & 0 \\
    SUSPICIOUS\_LINES\_FOUND\_OUT\_OF\_PR\_DIFF & 62 \\
    \rowcolor[rgb]{0.949,0.949,0.949}
    OUT\_OF\_SCOPE\_CI\_FAILURE & 98 \\
    FLACOCOBOT\_BUILD\_REPRODUCTION\_ERROR & 270 \\
    \rowcolor[rgb]{0.949,0.949,0.949}
    FAILURE\_HIDDEN\_BY\_FORCE\_PUSH & 18 \\ 
    PR\_NOT\_FAILED & 5 \\
    \rowcolor[rgb]{0.949,0.949,0.949}
    FLACOCOBOT\_CRASH & 4 \\ 
    \hline
    \textbf{Total} & \textbf{462} \\
    \bottomrule
    \end{tabular}
\end{table}

\paragraph{\texttt{SUSPICIOUS\_LINES\_FOUND\_IN\_PR\_DIFF}}
\flacocobot managed to find suspicious locations matching the pull request's diff in 5 out of 356 cases (1.40\%).
In particular, 4 PRs are related to INRIA Spoon project (\#4672, \#4702, \#4709, and \#4751) and 1 PR is related to Apache Dubbo project (\#10231).
Considering the status of projects at the beginning of the experiment (April, 4th, 2022), both of them have more than 2,000 Java files, summing over 150,000 lines of code. INRIA Spoon has more than 2,000 test cases, while Apache Dubbo has more than 3,000. Moreover, Apache Dubbo is a multi-module Maven project with 24 modules.
The projects also span across two different bytecode versions (Java 11 and Java 8).
These numbers show that, despite these projects being particularly complex, \flacocobot is able to perform fault localization on them and suggest suspicious lines to developers.


\paragraph{\texttt{SUSPICIOUS\_LINES\_FOUND\_OUT\_OF\_PR\_DIFF}}

\flacocobot found suspicious locations that are not contained in the changes introduced by the pull requests in 62 out of 356 cases (17.42\%).
For instance, considering pull request \#9893 of Apache Dubbo project, \flacocobot{} found 102 suspicious lines, but none of them belong to pull request's diff.
This is also a success for \flacocobot{}.
The fact that the lines cannot be output to the developer is related to a core limitation of the \github suggestion API we use. 

\paragraph{\texttt{OUT\_OF\_SCOPE\_CI\_FAILURE}}

In 98 cases, \flacocobot did not find suspicious locations because the failures associated with the pull requests are not managed by the fault localization tool.
This scenario happens, for example, with failures related to code quality errors (e.g., PR \#4699 of INRIA Spoon where CI reports errors generated by Qodana or PR \#3374 of Debezium where CI reports errors related to the commit message format), code coverage errors (e.g., PR \#4024 of Apache RocketMQ where CI reports errors generated by Coveralls), and errors present in the CI System (e.g., PR \#4010 of Apache RocketMQ where CI reports errors because no jobs were defined).
Therefore, for these cases, it is normal that \flacocobot was unable to find suspicious locations.

\paragraph{\texttt{FLACOCOBOT\_BUILD\_REPRODUCTION\_ERROR}}
In 270 out of 356 cases (75.84\%), \flacocobot had problems in building the projects related to:
\begin{inparaenum}[\it 1)]
    \item errors in building the project (e.g., errors in resolving the dependencies like with PR \#2208 of Vert-x3 Vertx Web),
    \item errors in retrieving all the test cases that are executed in the CI system (e.g., for PR \#2202 of Vert-x3 Vertx Web, \flacocobot did not find any test case, while the CI had 1,404 test cases and one test error, or
    \item errors in reproducing the original test failures (e.g., considering PR \#9898 of Apache Dubbo, \flacocobot executed the test cases, but the failure was not reproduced in the local environment). This confirms previous research \cite{urli2018}, showing that reproducing CI failures is challenging.
\end{inparaenum}

\paragraph{\texttt{FAILURE\_HIDDEN\_BY\_FORCE\_PUSH}}

In 18 out of 356 cases (5.05\%), \flacocobot detected a failure subsequently hidden by a force-push.
These are particular cases in which the developers decided to overcome the failure by forcing commits that override the failure-inducing changes, letting the CI pass again.
Debezium is the project for which \flacocobot encountered the highest number of these cases (15 out of 18).
Here, the suggestions from \flacocobot{} are not useful, as they were computed from versions that no longer exist in the version control system.

\paragraph{\texttt{FLACOCOBOT\_CRASH}}
In 4 out of 356 cases (1.12\%), \flacocobot had problems during its execution, making it impossible to report suspicious locations.
These happened with 3 PRs of Vert-x3/Vertx Web project (\#2161, \#2166, and \#2167), for which a timeout error prevented \flacocobot from ending the fault localization process.
The other pull request for which \flacocobot had problems is \#18173 of Apache Shardingsphere. In this case, there were errors during the instrumentation of the program that led to an \texttt{UnsupportedOperationException} exception.

\fbox{\begin{minipage}[t]{0.45\textwidth}
\textbf{Answer to RQ$_2$}: \flacocobot managed to find suspicious locations for at least one pull request in 9 out of 10 Java complex projects, showing its capability to perform fault localization in the field.
To the best of our knowledge, this is the largest ever fault localization experiment done on real failures in the field.
It validates the design and implementation of \flacoco, incl. doing binary code instrumentation with an industry-grade code instrumentation library, \jacoco{}.
\end{minipage}}

\section{Discussion}
\label{sec:discussion}

\subsection{Developer Opinion on \flacocobot's Concept}
\label{sec:developers}

We contacted the developers of 3 sampled projects by email. We asked permission to run \flacocobot in the project.
The contributors of three projects (Apache/Dubbo, Jenkinsci/Kubernetes Plugin, and INRIA/Spoon) gave us permission to run \flacocobot and post comments on their PRs.
This shows an indication that there is interest for a CI integrated fault localization service.
Of the developers who accepted to run \flacocobot, two developers did so because they considered it could help them locate suspicious causes when there are test failures in pull requests.
Another developer, from Jenkinsci/Kubernetes Plugin project, accepted and asked us about steps to perform from his side.


We collected feedback from the developers from INRIA/spoon involved in the failing PR \#4709.
This is a PR on which \flacocobot{} was able to leave a comment with fault localization suggestions.

The initial reaction from the developers was to react with a ``confused'' emoji on the comment left by \flacocobot.
When interviewed, the main point they remarked was related to the lack of explainability (i.e., how a score is computed) and interpretability (i.e., how good a score is) of the suspiciousness scores.
To our knowledge, this is a blind spot in fault localization research.



\subsection{Limitations}
Although \flacocobot has shown its capability to perform fault localization in the field with complex projects, there are interesting limitations that open space for future work.

\textbf{Test drivers}: 
Not all projects follow the same structure (e.g., complex projects have many submodules, each with test cases), or write tests under the same assumptions.
This aspect prevents \flacocobot from working to its full potential, as it does not detect all test cases and, therefore, is unable to compute the correct suspiciousness scores.
This shows that test harnesses have a real effect on fault localization. To our knowledge, \flacoco is the Java fault location tool that supports the largest number of test drivers.

\textbf{CI build filtering}:
CI build failures do not always happen due to test failures.
They can happen, for example, due to failures in the formatting or deployment steps.
\flacocobot{} is agnostic to the platform and configuration of each particular case and, as such, is not able to always filter out unwanted cases.
It is an open research question to perform static analysis of CI build scripts and logs to only consider the cases related to failed tests.

\textbf{Reproducibility of test failures}:
Tests that fail on CI do not always fail in the local environment where \flacocobot{} tries to reproduce the failure and performs fault localization.
To minimize this aspect, \flacocobot{} launches an isolated Docker environment but, sometimes (e.g., when the tests are not correctly detected, or when the test behavior depends on some specific configuration like the operating system), it is not enough to reproduce the exact same build.
There is a need for more research on exact CI failure reproduction.

\subsection{Future Improvements}
\label{sec:future_improvements}

We have found cases in which suspicious lines are not in the set of lines changed by the pull request.
As we have seen, this aspect can be related to the fact that the tests failing in \flacocobot's local environment are different from the tests failing on CI.
A possible way to address this problem would be to compare the failing tests detected by \flacocobot with the failing tests detected by CI.
When these two sets are different, \flacocobot would only consider the spectrum of the tests failing on the CI in the suspiciousness calculations.

An alternative to reproducing test failures in a local environment, and potentially avoid build reproduction errors, would be to provide \flacocobot{} as a CI step in common CI platforms (e.g., as a GitHub Action).
However, this option is dependant on developers actively integrating \flacocobot{} in their CI workflows.

Moreover, the developers could have control over this design option, and decide whether to receive fault localization suggestions for lines outside the diff or not.

Based on feedback from the interviewed developers, we believe it would be useful for a future version of \flacocobot{} to support developer specific configuration to, for example:
\begin{inparaenum}[\it 1)]
    \item set the suspiciousness score threshold,
    \item set a specific JDK and environment for build reproduction 
\end{inparaenum}

\section{Related work}
\label{sec:related-work}

\subsection{Spectrum-Based Fault Localization Tools}



In this section, we first focus on fault localization tools for Java code, as \flacoco{} targets that language, and then on tools for other languages.
The survey of fault localization tools presented by Archana et Agarwal~\cite{Archana2022SBFLP} discusses three tools for Java released in the last 10 years. We now describe them.
Campos et al.~\cite{gzoltar_eclipse} propose \gzoltar{}, a toolset for debugging programs.
The main difference between \gzoltar{} and \flacoco{} is that \gzoltar{} performs its own code instrumentation, while \flacoco{} relies on the worldwide used third-party library \jacoco{}.
This design decision allows \flacoco{} to be able to analyze the most recent versions of Java code, such as Java 17.
\gzoltar{} has been included in the implementations of automated program repair approaches such as Astor~\cite{martinez:hal-01321615}.
From the research conducted by Liu et al.~\cite{Liu2019FL} and Liu et al.~\cite{Liu2020EfficiencyAPR}, we observe that the version of \gzoltar{} most used in repair experiments is v0.1.1.
Nonetheless, versions up to v1.7.2 have been adopted, with Liu et al.~\cite{Liu2019FL} reporting an increase in the number of localized bugs by \gzoltar{} v1.6.0 on Defects4J compared with v0.1.1.

The recent work by Yang et al.~\cite{Yang2022BiasAPR}, which inspects bias on program repair evaluations, reports an extremely low efficiency of version 1.7.2 and uses version 0.1.1 instead.
As we have seen in this paper, \flacoco{} overcomes the issues related to these two versions of \gzoltar{} as it shows:
\begin{inparaenum}[1)]
    \item lower execution times than \gzoltar{} v1.7.3 (issue raised by \cite{Yang2022BiasAPR}),
    \item more support to Java LTS bytecode versions than \gzoltar{} v0.0.1.
\end{inparaenum}

Ribeiro et al.~\cite{Ribeiro2018SBFL} presented Jaguar, a spectrum-based fault localization tool for Java based on control-flow and data-flow spectra, also based on \jacoco{}.
\flacoco, on the other hand, is solely based on control-flow, and provides better integration via an API and AST bridge.
\flacoco also innovates by running tests in isolation from the main process, which allows for controlling classpaths, execution times as well as instrumentation at class loading time, and by supporting exception bugs, something that \jacoco{} does not support natively.
Furthermore, to our knowledge, Jaguar has only been evaluated in a subset of buggy versions of Defects4J (version 1).
In this paper, we show the ability of \flacoco{} to work, not only in Defects4J compiled to the most recent Java LTS version (Java 17), but also in the field.

Horváth et al.~\cite{Horvath2020} presented a Java tool, embedded as an Eclipse plug-in, that implements an interactive fault localization approach named iFL.
It presents suspicious lines computed using SBFL, one by one, to the developer, adjusting the suspiciousness scores based on her feedback.
It was evaluated on both artificial and Defects4J bugs, with students and professors as users.
The experiment we carried out in this paper was in the field: \flacoco{} was able to build and execute projects on real pull requests.

%



There are also tools that focus on other languages. 
For instance, for Python, 
Sarhan et al.~\cite{sarhan2021charmfl} present CharmFL\footnote{\url{https://sed-szeged.github.io/SpectrumBasedFaultLocalization/}}, an SBFL tool for Python, available as a PyCharm IDE plugin.
Other fault localization tools for Python are Fault-localization\footnote{\url{https://pypi.org/project/fault-localization/}} and Pinpoint\footnote{\url{https://pypi.org/project/pytest-pinpoint/}}.
For SBFL tools for C, Janssen et al.~\cite{zoltar} present a tool called Zoltar\footnote{\url{https://github.com/ncfxy/zoltar}}, Golagha et al. present Aletheia~\cite{Golagha2018Aletheia}, a failure diagnosis toolchain, which also performs fault localization. Both tools are based on the technique BARINEL~\cite{barinel}.

Fault localization tools have also been integrated into existing IDEs such as 
Eclipse (e.g. \cite{gzoltar_eclipse}), PyCharm (e.g. \cite{sarhan2021charmfl}) and Visual Studio (e.g. \cite{flavs}, \cite{leao2022integration} which uses \flacoco{}).

\subsection{Other FL Approaches and Tools}

Papadakis and Le Traon~\cite{Papadakis2015Metallaxis} propose Metallaxis, a technique based on mutation analysis.
Zeng et al. present SmartFL~\cite{Zeng2022SmartFL} a fault localization approach based on probabilistic graph model.
Li et al.~\cite{Li2021FLImage} propose DeepRL4FL, a deep
learning fault localization approach based on image pattern recognition problem.
Kucuk et al.~\cite{Kucuk2021@improvingFlPredicate} propose UniVal, a technique that is both coverage-based and value-based, which uses causal inference techniques and machine learning to integrate information about both predicate outcomes and variable values.
Li et al.~\cite{Li2019DeepFL} present DeepFL, a deep learning-based approach to predict potential fault locations incorporating various dimensions of fault diagnosis information.
Zou et al.~\cite{Zou2021CombineFL} evaluate different techniques, including spectrum-based fault localization, mutation-based fault localization, and dynamic slicing, and propose an approach named CombineFL that combines these techniques.
Lou et al.\cite{Lou2021} propose Grace, coverage-based fault localization technique, which fully exploits coverage information via graph-based representation learning. 

For all these, still recent, contributions, it is unknown how the proposed techniques will perform in the wild.
We also note that the implementations of these approaches are either not publicly available or were built in order to be exclusively executed in controlled environments such as Defects4J (e.g., via a virtual machine).
To our knowledge, our fault localization results, targeting real pull requests, have significantly higher external validity than related research.

\subsection{Software Bots}

Different studies have presented software engineering bots that interact with developers by writing comments on PRs~\cite{Santhanam2022BotsMapping}.
For example, Khanan et al.~\cite{Khanan2022JustInTime} design a defect prediction bot to compute the riskiness of each commit and present it on a PR's comment.
\flacocobot{} has a different goal: to pinpoint the locations that potentially have the failure-inducing bug.

Other works propose bots for code refactorings via \prs{}~\cite{Wyrich2020PerceptionAA}. 
For example, 
RefBot~\cite{Alizadeh2019RefBot}) is designed for code smell removal,
RefDoc~\cite{Rebai2019RefDoc} is designed for refactoring documentation,
CCBot~\cite{Carr2017CCBot} is meant to add contracts to detect faults, 
FixMe~\cite{Phaithoon2021FixMe} is a \github bot for detecting and monitoring self-admitted technical debt.
The goal of \flacocobot{} is to enrich existing PRs with new information to help developers with the debugging task, rather than creating a new PR.

Ochoa et al.~\cite{Ochoa2022BreakBot} present BreakBot, a bot that performs static analysis on \prs{} from Java libraries on \github to identify the breaking changes they introduce and their impact on client projects.
BreakBot and \flacocobot{} are complementary: the former focuses on changes that break users (clients) of the program under change, while \flacocobot{} on changes in a project that break its expected behavior.

All these bots are static in essence (i.e., they parse and transform code but do not execute it), as opposed to \flacocobot{} which performs a dynamic analysis of the target projects.
Dynamic analysis is arguably more challenging because it requires build scripts and appropriate execution environments.



\section{Conclusion}
\label{sec:conclusion}

In this paper, we presented \flacoco{}, a fault localization tool for Java based on \jacoco{}, an industry-grade code coverage library.
\flacoco{} is able to produce fault localization results for different versions of Java bytecode, including the most recent one (Java 17).
To show the ability of \flacoco{} to work on complex projects, we designed and implemented an integrated version of~\flacoco{} on \github, which proposes fault localization results on failing pull requests from \github.
\flacocobot{} monitored 10 complex open-source Java projects and was able to find fault localization results in 9 projects.
\flacoco{} is made publicly available for both researchers and practitioners at \url{https://github.com/SpoonLabs/flacoco}.

\balance
\bibliographystyle{IEEEtran}
\bibliography{IEEEabrv,main}

\clearpage





\end{document}